\documentclass{article}

\PassOptionsToPackage{square,numbers,sort&compress}{natbib}

\usepackage{authblk}

\usepackage[utf8]{inputenc} % allow utf-8 input
\usepackage[T1]{fontenc}    % use 8-bit T1 fonts
\usepackage[hidelinks]{hyperref}       % hyperlinks
\usepackage{url}            % simple URL typesetting
\usepackage{booktabs}       % professional-quality tables
\usepackage{amsfonts}       % blackboard math symbols
\usepackage{bm}
\usepackage{nicefrac}       % compact symbols for 1/2, etc.
\usepackage{microtype}      % microtypography
\usepackage{xcolor}         % colors
\usepackage{bbm}
\usepackage{dsfont}
\usepackage[
  separate-uncertainty = true,
  multi-part-units = repeat
]{siunitx}
\usepackage{fixmath}
\usepackage[top=2cm,bottom=2cm,left=3cm,right=2.5cm]{geometry}

\usepackage{graphicx}
\usepackage{amsmath,mathtools,amsthm}
\usepackage{enumitem}
\graphicspath{{figs/}}
\usepackage{wrapfig}
\usepackage{algorithm}
\usepackage{algpseudocode}
\usepackage{caption}
\usepackage{subcaption}
\usepackage{comment}
\usepackage{xr}

\usepackage{xcolor}
\definecolor{linkcolor}{HTML}{0000FF}
\hypersetup{colorlinks=true, linkcolor=linkcolor}

\makeatletter
\newcommand*{\addFileDependency}[1]{
  \typeout{(#1)}
  \@addtofilelist{#1}
  \IfFileExists{#1}{}{\typeout{No file #1.}}
}
\makeatother

\usepackage{booktabs}

\usepackage{mathtools}
\mathtoolsset{showonlyrefs}

\DeclareMathOperator*{\argmin}{arg\,min}

\usepackage{bbm}

\def\vx{{\bm{x}}}

\def\mQ{{\bm{Q}}}

\title{Lightsolver challenges a leading deep learning solver for Max-2-SAT problems}

\author[1]{Hod~Wirzberger}
\author[1]{Assaf~Kalinski}
\author[1]{Idan~Meirzada}
\author[1]{Harel~Primack}
\author[1,2]{Yaniv~Romano}
\author[1]{Chene~Tradonsky}
\author[1]{Ruti~Ben~Shlomi}

\affil[1]{LightSolver LTD., Tel Aviv, Israel}
\affil[2]{Departments of Electrical and Computer Engineering and Computer Science, Technion IIT, Israel}

\begin{document}

\date{}

\maketitle

\begin{abstract}
Maximum 2-satisfiability (MAX-2-SAT) is a type of combinatorial decision problem that is known to be NP-hard. In this paper, we compare LightSolver's quantum-inspired algorithm to a leading deep-learning solver for the MAX-2-SAT problem. Experiments on benchmark data sets show that LightSolver achieves significantly smaller time-to-optimal-solution compared to a state-of-the-art deep-learning algorithm, where the gain in performance tends to increase with the problem size. 
\end{abstract}

\section{Introduction}

A wide range of real-world problems with substantial societal, economic, and scientific implications can be posed as combinatorial optimization tasks. Advances in combinatorial optimization have led to more efficient transportation systems, supply chains, resource management, and more \cite{paschos2014applications,naseri2020application,fortun1993scientists,magnanti1981combinatorial,greenberg2004opportunities}. In this work, we consider the classic maximum 2-satisfiability (\texttt{MAX-2-SAT}) problem~\cite{mezard2009information}, which is ubiquitous in scheduling or resource allocation tasks, to name a few applications~\cite{du1997satisfiability}. 

Suppose we are given a set of $N$ binary variables $\vx = (\vx_1, \vx_2, \dots, \vx_N)$ and a set of $C$ constraints (or clauses) with two variables per clause that form a Boolean formula $F(\vx)$. Our goal is to assign a binary value to each variable $\vx_i$ such that the maximal number of the clauses are satisfied. The Boolean formula $F(\vx)$ we consider takes a conjunctive normal form, consisting of a conjunction (logical AND) of clauses, where each clause is a disjunction (logical OR) of literals. For example, the formula
\begin{equation}
\label{eq:example}
    F(\vx) = (\vx_1 \lor \vx_2 ) \land (\neg \vx_1 \lor \vx_2 )  \land ( \vx_1 \lor \neg \vx_2 ) \land ( \neg \vx_1 \lor \neg \vx_2 )
\end{equation}
has $N=2$ variables and $C=4$ clauses. The notations $\land$, $\lor$, and $\neg$ stand for the logical AND, OR, and negation operators, respectively. While there is no assignment of $\vx_1$ and $\vx_2$ that satisfy the specific formula given above, we can make 3 out of 4 clauses true by assigning $\vx_1 = 1$ and $\vx_2 = 1$. Hence, the optimal \texttt{MAX-2-SAT} solution for this example is 3 clauses.

The \texttt{MAX-2-SAT} problem is known to be NP-hard~\cite{arora2009computational}, implying that exact solutions can be merely obtained for relatively small problems; the computational complexity rapidly increases with the number of variables \cite{khot2007optimal,khanna2001approximability}. Yet, there exist computationally efficient approximation algorithms that are capable of finding useful solutions that are not necessarily the optimal ones. More traditional methods for tackling constraint satisfaction problems include combinatorial constraint propagation algorithms, logic programming techniques, and semi-definite programming \cite{williams2005new,luo2014ccls,larrosa2008logical,khot2007optimal,raghavendra2008optimal}. A different line of work harnesses modern machine and deep learning algorithms, such as graph neural networks, to achieve high-quality solutions in amendable time complexity~\cite{toenshoff2021graph,lemos2019graph,prates2019learning,selsam2018learning,li2018combinatorial}, leveraging the impressive computing power of graphical processing units (GPUs). Another approach to tackle the \texttt{MAX-2-SAT} problem is via quantum or quantum-inspired computing platforms \cite{santra2014max,bravyi2011efficient,mirkarimi2022comparing}. A typical approach here is to formulate \texttt{MAX-2-SAT} as a quadratic unconstrained binary optimization (QUBO) problem \cite{kochenberger2005using,glover2018tutorial}, which, in principle, can be efficiently solved on quantum computers, annealers, or simulators~\cite{bian2020solving,oshiyama2022benchmark,mirkarimi2022comparing}.

In this paper, we compare the performance of a leading off-the-shelf deep learning solver for the \texttt{MAX-2-SAT} problem, called RUN-CSP~\cite{toenshoff2021graph}, to the QUBO solver of \hyperlink{www.lightsolver.com}{LightSolver}~\cite{meirzada2022lightsolver, Romano_quantum_sparse_coding}: a quantum-inspired digital computing platform that simulates its all-optical devices. In our experiments, we make two important design choices to ensure a fair comparison. First, both RUN-CSP and LightSolver algorithms are executed on the very same GPU computing resources. Second, we use the \texttt{MAX-2-SAT} benchmark data sets included in the official software package of RUN-CSP~\cite{toenshoff2021graph}.\footnote{Code is available online at \url{https://github.com/RUNCSP/RUN-CSP}} This experimental protocol allows us to apply RUN-CSP with the training scheme and set of hyper-parameters that the authors chose to achieve the best performance. Importantly, we evaluate the performance of the two methods by comparing their time-to-exact-solution, where the runtime does not include the training phase of RUN-CSP or the preprocessing required to formulate the QUBO problem in the case of LightSolver's algorithm. Our experiments reveal that (i) the average time-to-exact-solution obtained by LightSolver's algorithm is 2X to 1000X faster than that of RUN-CSP across all the data sets examined; and (ii) the gain in performance tends to increase with the size of the problem, which is defined by the number of variables $N$.

\section{Methods}

\subsection{Benchmark: RUN-CSP}
The RUN-CSP algorithm \cite{toenshoff2021graph} is a generic deep-learning-based solver for maximum binary constraint satisfaction problems. At a high level, the idea is to train a recurrent graph neural network on a set of instances of the constraint satisfaction problem of interest by minimizing a loss function that rewards solutions that satisfy a large number of constraints. The update of the network parameters can be thought of as a message-passing algorithm. Importantly, once the training is complete, the fitted model can be applied to new test instances of arbitrary size as the network parameters are shared across all variables. Interestingly, RUN-CSP was shown to outperform state-of-the-art heuristic algorithms for solving the \texttt{MAX-2-SAT} problem, motivating our desire to compare LightSolver's algorithm to RUN-CSP.

\subsection{Proposed: LightSolver}

To solve the \texttt{MAX-2-SAT} problem on LightSolver's platform we formulate this task as a QUBO problem, which, in general, is given by~\cite{glover2018tutorial}
\begin{equation}
\label{eq:qubo}
    \min_{\vx \in \{0,1\}^N} \vx^T \mQ \vx.
\end{equation}
Above, $\mQ \in \mathbb{R}^{N \times N}$ is a QUBO matrix and $\vx$ is the vector of binary unknowns---the literals. To translate the Boolean function of interest $F(\vx)$ into the above quadratic form, we observe that, in general, there are three possible clause types that should be mapped to their corresponding quadratic forms, as summarized below~\cite{glover2018tutorial}.
\begin{center}
\begin{tabular}{l|l}
    Clause & Quadradic Mapping \\ \hline 
    $\vx_i \lor \vx_j$ & $1-\vx_i -\vx_j + \vx_i \vx_j$  \\
     $\neg\vx_i \lor \vx_j$ & $\vx_i - \vx_i \vx_j$  \\
    $\neg \vx_i \lor  \neg \vx_j$ & $\vx_i \vx_j$ 
\end{tabular}
\end{center}
Notably, the mapping is defined such that when a clause is satisfied the value of the quadratic term is 0. By contrast,  when the clause is not met the quadratic mapping is equal to 1. Lastly, we sum the quadratic mappings of all clause terms and get the final QUBO problem. In our running example, the formula \eqref{eq:example} can be mapped into the following objective:
\begin{equation}
    \min_{\vx_1, \vx_2 \in \{0,1\}} (1-\vx_1 -\vx_2 + \vx_1 \vx_2) + (\vx_1 - \vx_1\vx_2)  + (\vx_2 - \vx_2\vx_1) +  (\vx_1\vx_2).
\end{equation}
Observe that $\vx_i = \vx_i \vx_i $ since $\vx_i$ is binary, and therefore the above objective perfectly fits into \eqref{eq:qubo}---it is nothing but a sum of quadratic elements. One attractive feature of the above formulation is that the matrix $\mQ$ is of size $N \times N$, regardless of the number of clauses~\cite{glover2018tutorial}.

\subsection{Experiments}

We follow \cite{toenshoff2021graph} and compare RUN-CSP to LightSolver on the benchmark instances from the unweighted track of the Max-Sat Evaluation 2016~\cite{bacchus2020maxsat}, for which the optimal solution of each problem is known. The number of variables and clauses of the different benchmark instances are summarized in Table~\ref{tab:benchmarks}. 
\begin{table}[h!]
\centering
\begin{tabular}{c|c|c}
    Instance name & Variables $N$ & Clauses $C$\\ \hline 
    t3pm3 & 27 & 162 \\
    t4pm3 & 64 & 384 \\
    t5pm3 & 125 & 750 \\
    t6pm3 & 216 & 1269 \\
    t7pm3 & 343 & 2058 \\
\end{tabular}
\caption{Properties of the benchmark instances of the unweighted track of the Max-Sat Evaluation 2016.}
\label{tab:benchmarks}
\end{table}

The evaluation metric we used is `time-to-optimal-solution' defined as~\cite{meirzada2022lightsolver,kowalsky20223}
\begin{equation}
    \text{Time-to-optimal-solution} = \argmin_{t>0} \ t \cdot \frac{\ln(1-0.99)}{\ln(1-p_i(t))},
\end{equation}
where $t$ is the simulation time, and $p_i(t)$ is the empirical probability of reaching an optimal solution. In plain words, we report the smallest simulation time required to obtain an optimal solution at least once, with 99\% probability.
For RUN-CSP, we swept over a set of possible hyper-parameters and reported the choice that achieved the best time-to-optimal-solution. 

\begin{figure}[H]
    \centering
    \includegraphics[width=0.75\textwidth]{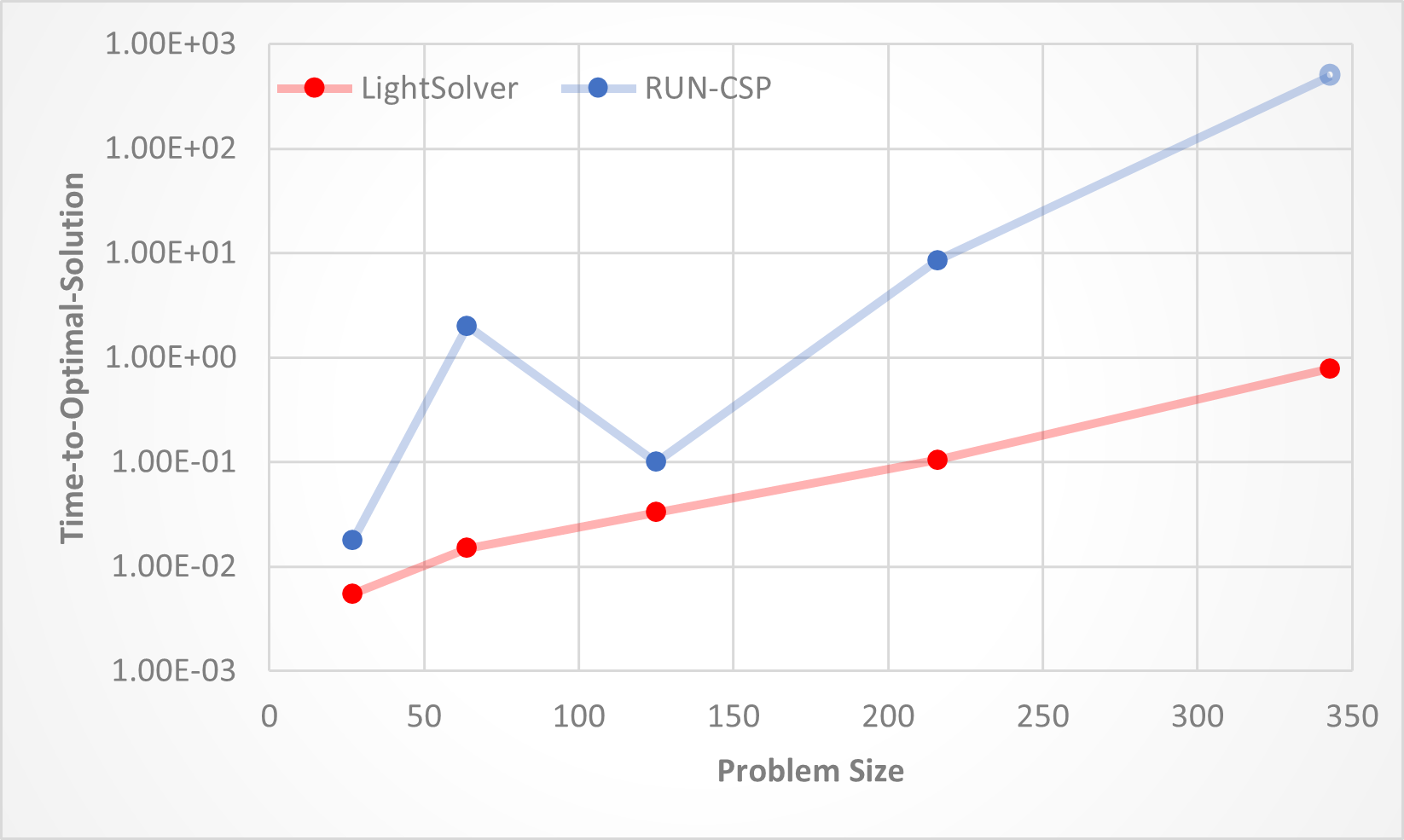}
     % \hspace{0.1cm}
     %\hfill
    \caption{Comparison between the time-to-optimal-solution of LightSolver and RUN-CSP algorithms, evaluated on benchmark \texttt{MAX-2-SAT} problems.}
    \label{TTS}
\end{figure}

The results are summarized in Figure~\ref{TTS}, where both RUN-CSP and LightSolver are evaluated on NVIDIA T4 GPU on the AWS platform. As portrayed, LightSolver consistently achieves the optimal solution, with significantly smaller execution times for larger problem sizes. For the largest problem, RUN-CSP did not manage to reach the optimal solution after 400 seconds, converging to the first excited state at best (white marker). 
\\

\section*{Conclusions}

In this work, we compared LightSolver quantum-inspired algorithm to a leading deep learning approach for solving the NP-hard \texttt{MAX-2-SAT} problem. 
Deep learning networks are a popular strategy for facing complex optimization problems due to their wide application potential. However, they require a costly training step prior to providing solutions. 
Quantum and quantum-inspired solvers, on the other hand, do not require a training stage and thus can function as an on-demand resource. Our experiments show that the LightSolver simulator achieves 2X-1000X faster time-to-optimal-solution for solving the Max-2-SAT problems compared to the deep learning solver, for problems of small to medium scale. 
This result further supports the immense potential of quantum computers and quantum-inspired platforms to yield better solutions for challenging combinatorial optimization problems. 

\bibliographystyle{unsrt}
\bibliography{refs}

\end{document}